# Ultrafast terahertz responses in monolayer graphene

Q. Jin[1], H.M. Dong[1*], K. Han[1] and X.F. Wang[2]

[1] Department of Physics, China University of Mining and Technology, Xuzhou 221116, China
[2] College of Physics, Optoelectronics and Energy, Soochow University, Suzhou 215006, China
*E-mail: hmdong@cumt.edu.cn

**Abstract**. We theoretically investigate the ultrafast terahertz(THz) properties of monolayer graphene. The analytical formulations of the photon carrier, electric polarization and optical current are obtained by solving the Bloch-equations in present of the ultrafast THz Gaussian pulse. Graphene shows a large nonlinear and ultrafast optical response at THz frequencies due to the gapless and relativistic Dirac particles with nearly linear energy dispersion. It is found that the photon carrier density, electric polarization and optical current density increase with increasing the frequency of the THz pulse. These theoretical results are in agreement with recent experimental findings. This study confirms further that graphene exhibits important features and is relevant to the applications in the ultrafast THz fields.

## 1. Introduction

The graphene, a single layer of carbon atoms enmeshed in a honeycomb-like, exhibits the unique optical and electronic properties[1], such as quantum hall effect[2,3], high mobility[4], and the universal optical conductance[5]. Moreover, the experimental results using four-wave mixing demonstrate that graphene exhibits a very strong nonlinear optical response in the near-infrared spectral region[6]. Recently, ultrafast optical properties of monolayer graphene are widely researched in experiments, such as the utrafast dynamic electron relaxation, saturation nonlinear optical property, and the ultrafast mode-locked. It shows that graphene play an important application in ultrafast mode-locked laser, optical modulator, terahertz optical detection, soliton and ultrafast optical communication. Due to Pauli exclusion-principle, graphene possesses the characteristics of ultrafast saturable absorber which can produce ultrafast mode-locked laser pulse[7]. Ruzicka and co-workers have studied the expansion of a Gaussian spatial profile of carriers excited by a 1761 nm probe pulse and observe the expansion of the carrier density profile decreases to a slow rate within 1 ps[8]. Dawlaty[9] has measured carrier relaxation times of epitaxial graphene layers which grown on SiC wafers with the ultrafast optical pump-probe spectroscopy, and found the fast relaxation transient range between 70~120 fs. Liu studied the ultrafast relaxation dynamics and nonlinear optical response in single and few-layered graphene. It is found that the carrier dynamics of graphene meet a bi-exponential decay with a few hundreds of femtoseconds [10]. It has been found that the ultrafast saturable absorption of graphene is in the femtosecond (fs) time experimentally[11]. The results also reveal that graphene possesses great potential for applications as passive mode-lockers, optical pulse shapers or output couplers. D. Brida exploited the temporal resolution of the pump-probe scheme to observe the relaxation dynamics of out-of-equilibrium electrons in single layer graphene[12]. They have found that the ultrafast timescale of this process is strongly related to the particular band structure and the electronic properties of graphene. D. Sun has investigated the effect of hot carriers on a coherently injected current in graphene and found the THz emission from the coherently controlled ballistic current is suppressed when a 100 fs pulse[13]. The above interesting and important experimental studies show that graphene exhibits important features and applications in ultrafast dynamic optical field and can be applied for some areas of ultrafast THz technology. However, we have noticed that the relevant theoretical researches are relatively absent. In addition, we have investigated the steady-state optical properties of the graphene [5] and the dynamic optical properties of the doped graphene [14,15]. Motivated by the experimental works, we would establish the theory model to study the ultrafast THz responses of graphene to understand the dynamic THz properties and explain the experimental results.

## 2. Theoretical approach

In conjunction with recent experimental achievement in graphene, here we would like to develop a simple theoretical approach to study the ultrafast THz responses in monolayer graphene. The electrons of graphene move like Fermi-Dirac particles in the low energy regime. Using the effective-mass approximation, a carrier(electron or hole) in a monolayer graphene can be described by Weyl's equation for a massless Dirac particles. The corresponding eigenvalue is $E_\lambda(\mathbf{k}) = \lambda\gamma|\mathbf{k}|$, where $\gamma = \hbar v_F$, $v_F = 10^8 \text{cm/s}$ being the Fermi velocity of a Dirac particle in graphene. $\mathbf{k} = (k_x, k_y)$ is the wave vector for a carries along the graphene sheet. $\lambda = +1$ is for an electron in the conduction bands, and $\lambda = -1$ is for a hole in the valence bands in graphene. Correspondingly, the wave functions of graphene is that $\psi(\mathbf{r}) = |k\rangle = 2^{1/2}[1, \lambda e^{i\phi}]e^{i\mathbf{k}\mathbf{r}}$, $\phi$ is the angel between $\mathbf{k}$ and the x direction and $\mathbf{r}=(x,y)$.

We consider an air/graphene/dielectric-wafer system. A ultrafast THz pulse is applied perpendicular to the graphene layer and is polarized linearly along the x-direction of the graphene sheet. For a graphene in the presence of a light field, the system Hamiltonian can be written as: H=$H_c$+$H_{co}$, where $H_c$ and $H_{co}$ are, respectively, the single carrier Hamiltonian and carrier–photon Hamiltonian. According to quantum theory, the Hamiltonian of the two energy levels are

$$H_c = E_+(k)\hat{a}_{c,k}^+ \hat{a}_{c,k} + E_-(k)\hat{a}_{v,k}^+ \hat{a}_{v,k} \qquad (1)$$

$$H_{co} = -d_{cv}E(t)(\hat{a}_{c,k}^+ \hat{a}_{v,k}^+ + \text{h.c}) \ , \qquad (2)$$

Where $\hat{a}_{c,k}^+$ and $\hat{a}_{c,k}$ are respectively the creation and annihilation operators of the electrons in conduction bands, $\hat{a}_{v,k}^+$ and $\hat{a}_{v,k}$ are respectively the creation and annihilation operators of the holes in valence bands in a monolayer graphene. $E(t) = E_0 e^{-\omega^2 t^2/a} e^{-i\omega t}$ is the ultrafast THz pulse with $E_0$ being the intensity of the incident field and $\omega$ being the frequency of the photons. a=4ln2 and t is the time. Moreover, $d_{cv} = ev_F \sin\phi/\omega$ is the inter-band optical dipole matrix. According to Heisenberg equation, the Bloch equation of graphene can be written as

$$\frac{\partial p_k(t)}{\partial t} = -2iv_F k p_k(t) - i\Omega(t)(n_{e,k} + n_{h,k} - 1)$$

$$\frac{\partial n_{e,k}(t)}{\partial t} = -2\text{Im}[\Omega(t)p_k^*] \qquad (3)$$

$$\frac{\partial n_{h,k}(t)}{\partial t} = -2\text{Im}[\Omega(t)p_k^*].$$

Here $n_{e,k}(t)$ and $n_{h,k}(t)$ are respectively the distribution function of electron and hole densities for the $\mathbf{k}$ band. $p_k(t)$ is the inter-band microscopic polarization and $p_k^*(t)$ is the complex conjugate. The Rabi frequency of graphene is

$$\Omega(t) = \frac{d_{cv}E(t)}{\hbar} = \frac{ev_F E_0}{\hbar\omega}\sin\phi \, e^{-\omega^2 t^2/a}e^{-i\omega t} \ . \qquad (4)$$

With the Bloch equations (3), we can obtain $n_{e,k}(t)$, $n_{h,k}(t)$ and $p_k(t)$ for a given THz pulse E(t). Meanwhile, the photon induced carrier density can be obtained by $N_\lambda(t) = g_s g_v \sum_k n_{\lambda,k}(t)$, where $g_s$=2 and $g_v$=2 account, respectively, for spin and valley degeneracy. The total mesoscopic polarizability of graphene $P(t) = \text{Re}[g_s g_v \sum_k d_{cv}^* p_k(t)]$ in the presence of a radiation field.

In this paper we consider the intrinsic(undoped) graphene, in other word the valence band is fully occupied by electrons and the conduction band is empty in absence of the light fields. With rotating-wave approximation, we can obtain the photon-induced electron density N(t) in the presence of the weak THz pulse, which is

$$N(\tau) = \frac{\omega^2}{8\pi v_F^2}\left(1 - e^{\alpha\varepsilon(-2^{\tau^2})}J_b[0,\alpha\varepsilon(-2^{\tau^2})]\right). \quad (5)$$

In order to display physics clearly, here $\alpha = ae^2\gamma^2 E_0^2/(2\hbar^4\omega^4)$, $\tau = \omega t$ is the normalized time, $\varepsilon(x)$ is the exponential integral function. $J_b[n,x]$ is the first kind of modified Bessel function. The total polarizability of graphene can be expressed as

$$P(\tau) = -\frac{ae^2 E_0}{16\pi\hbar\omega}\frac{\sin(\sqrt{a}\tau)}{\sqrt{a}\tau}e^{\alpha\varepsilon(-2^{\tau^2})-\tau^2}\left(J_b[0,\alpha\varepsilon(-2^{\tau^2})] - J_b[1,\alpha\varepsilon(-2^{\tau^2})]\right). \quad (6)$$

Then, the photon induced optical current density $J(\tau)$ can be obtained by taking the derivative of polarizability versus time, which is $J(\tau) = dP(\tau)/d\tau$.

It is clear that the analytical formulations of the photon carrier, electric polarization and optical current are obtained by solving the Bloch-equations in present of the ultrafast THz Gaussian pulse. The analytical formulations are very useful and accurate, which can help us to understand the ultrafast THz responses in monolayer graphene.

## 3. Results and discussions

We have calculated the total electric polarization P($\tau$), photo induced current density J($\tau$) and photon induced electron density N($\tau$) over time with different frequencies $f$ ($\omega = 2\pi f$) in THz regain. The intensity of THz field $E_0$=10V/cm is a weak field. These conditions are consistent with the experiment[6,9].

In figure 1, we show the temporal evolution of photon induced electron density N($\tau$) for different frequencies of the THz pulse. It is shown that N($\tau$) obviously increases with increasing the incident frequency(energy) of the THz pulse. There are very high and effective photon induced carries in graphene in presence of the THz pulse. The photon induced electron density N($\tau$) can be up to $10^8$ cm$^{-2}$ within a THz optical filed. It shows that the carriers response to the THz pulse very sensitively. It exhibits an ultrafast response in terahertz regime, and the shorter the response time, the higher the THz frequency the response time. The physical reason is that the monolayer graphene is a zero-band system with linear energy dispersion. The Dirac quasi-particles can rapidly response to the THz photons in such system.

In figure 2, we show the electric polarization P($\tau$) over the time for the different frequencies f of the light field. The electronic polarization shows the strength of the material responding to external field. Electrons in the valence band can transit to the conduction band by absorbing the photons. Consequently, the material is electric polarized. It shows that the larger the photon energy, the electric the stronger of the polarization in graphene system. The electric polarization is unsynchronized with light field. The study indicates that graphene is a new nonlinear optical material. The graphene shows distinct nonlinear THz effect, even in weak light field. This make clear the graphene holds an important research and application value in THz non-linear optical areas. The results are consistent with the experimental results[6,10].

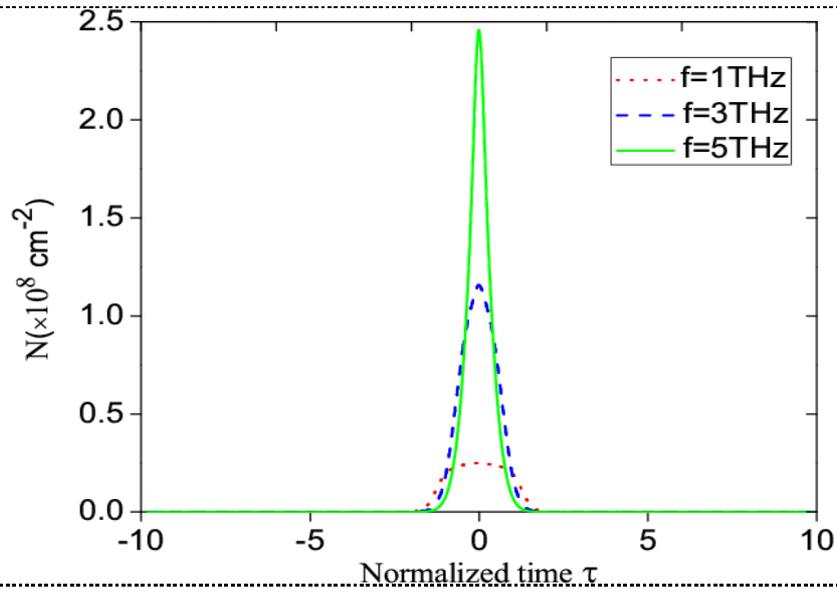

**Figure 1.** Photon-induced carriers N over normalized time $\tau$ with the different frequency of the ultrafast THz pulse.

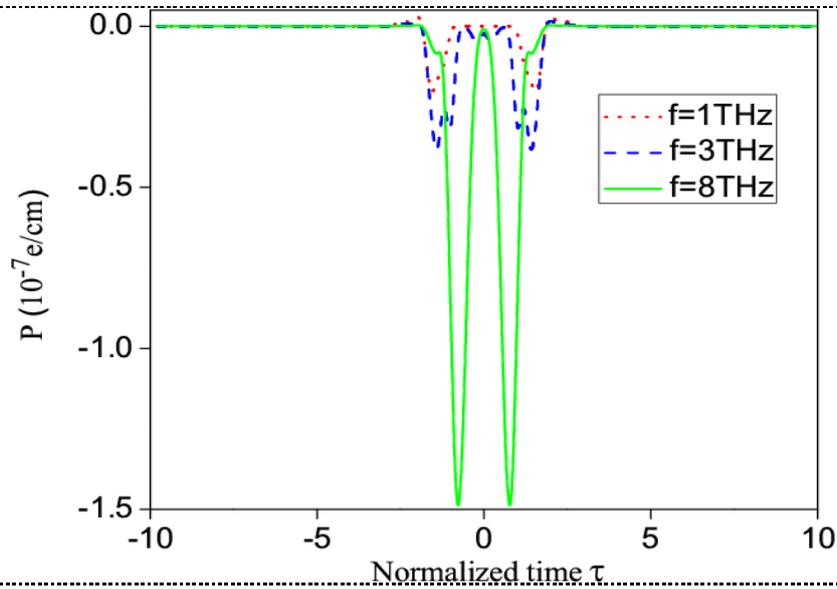

**Figure 2.** The electric polarization P over time with normalized time $\tau$ with the different frequency of the ultrafast THz pulse.

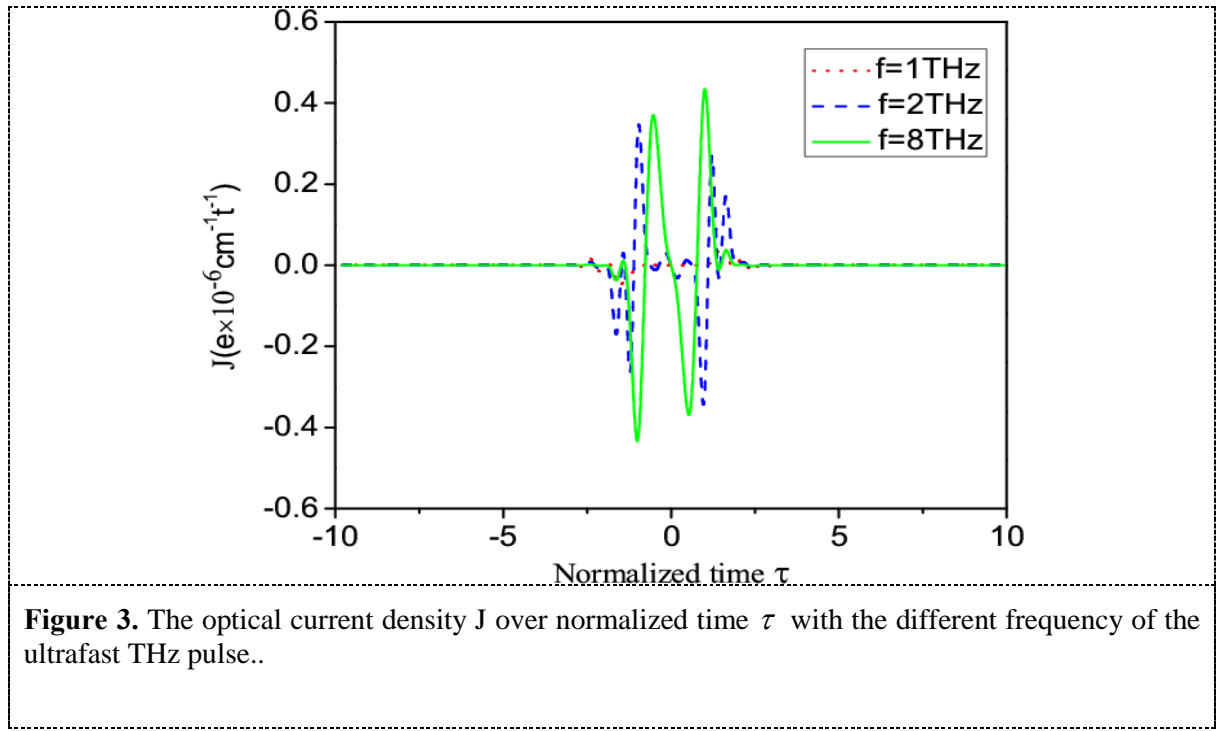

**Figure 3.** The optical current density J over normalized time $\tau$ with the different frequency of the ultrafast THz pulse..

In figure 3, the optical current density $J(\tau)$ over time is shown for the different THz frequencies. The optical current density is related to the electric polarization, which is $J(\tau) = dP(\tau)/d\tau$. The non-equilibrium carriers form by absorbing photons, which results in the electric polarization, and then leads to the optical current. It shows that the optical current density $J(\tau)$ becomes large with increasing the photon energy(frequency). The reason is that the more electrons transiting from the valence bands to the conduction bands, the larger the photon energy $\hbar\omega$. We find the optical current density $J(\tau)$ is not synchronized with THz pulse E(t). This is due to Dirac Fermi particles with the linear energy dispersion, which are very different with that in the conventional semiconductor material(CSM) and two dimensional electron gas(2DEG). It is clear that there is an ultrafast THz response in an intrinsic graphene system. This is very useful and important because the optical current can be built more rapidly than that in CSM and 2DEG.

## 4. Conclusions
In this paper, we investigate the ultrafast THz responses of graphene systems. The theoretical result shows that graphene has the distinct nonlinear ultrafast optical properties in terahertz regime. The photon carrier density, electric polarization and optical current density increase with increasing the frequency of the THz pulse. Graphene shows a large nonlinear and ultrafast optical response at THz frequencies due to the gapless and relativistic Dirac particles with nearly linear energy dispersion. This study shows that graphene is a new nonlinear ultrafast THz material. Our theoretical results can explain the results in the ultrafast optical experiments at low temperatures. This study is relevant to the application of graphene as THz nano-optical devices such as ultrafast sensitive THz devices.

**Acknowledges**
This work was supported by the Fundamental Research Funds for the Central Universities (No. 2015XKMS077).